\documentclass[aps,prl,reprint,amsmath,amssymb,superscriptaddress,showpacs]{revtex4-1}
\usepackage{bm}
\usepackage{graphicx}
\usepackage{subfigure}
\usepackage{color}
\usepackage{dcolumn}
\usepackage{kotex}
\usepackage{hyperref}

\begin{document}

\title{Magnetic Excitations of the Cu$^{2+}$ Quantum Spin Chain in Sr$_3$CuPtO$_6$ }

\author{J. C. Leiner}
\email{jleiner@snu.ac.kr}
\affiliation{Center for Correlated Electron Systems, Institute for Basic Science (IBS), Seoul 08826, Korea}
\affiliation{Department of Physics and Astronomy, Seoul National University, Seoul 08826, Korea}

\author{Joosung Oh}
\affiliation{Center for Correlated Electron Systems, Institute for Basic Science (IBS), Seoul 08826, Korea}
\affiliation{Department of Physics and Astronomy, Seoul National University, Seoul 08826, Korea}

\author{A. I. Kolesnikov}
\affiliation{Neutron Scattering Division, Oak Ridge National Laboratory, Oak Ridge, TN 37831, USA}

\author{M. B. Stone}
\affiliation{Neutron Scattering Division, Oak Ridge National Laboratory, Oak Ridge, TN 37831, USA}

\author{Manh Duc Le}
\affiliation{ISIS Facility, STFC, Rutherford Appleton Laboratory, Didcot, Oxfordshire OX11-0QX, United Kingdom}

\author{E. P. Kenny}
\author{B. J. Powell}
\affiliation{School of Mathematics and Physics, The University of Queensland, Brisbane, Queensland 4072, Australia}

\author{M. Mourigal}
\affiliation{School of Physics, Georgia Institute of Technology, Atlanta, GA 30332, USA}

\author{E. E. Gordon}
\author{M.-H. Whangbo}
\affiliation{Department of Chemistry, North Carolina State University, Raleigh, North Carolina 27695-8204, USA}

\author{J.-W. Kim}
\author{S.-W. Cheong}
\affiliation{Rutgers Center for Emergent Materials and Department of Physics and Astronomy, Rutgers University, Piscataway, New Jersey 08854, USA}

\author{Je-Geun Park}
\email{jgpark10@snu.ac.kr}
\affiliation{Center for Correlated Electron Systems, Institute for Basic Science (IBS), Seoul 08826, Korea}
\affiliation{Department of Physics and Astronomy, Seoul National University, Seoul 08826, Korea}

\date{\today}

\begin{abstract}
We report the magnetic excitation spectrum as measured by inelastic neutron scattering for a polycrystalline sample of Sr$_3$CuPtO$_6$. Modeling the data by the 2+4 spinon contributions to the dynamical susceptibility within the chains and with interchain coupling treated in the random phase approximation accounts for the major features of the powder averaged structure factor. The magnetic excitations broaden considerably as temperature is raised, persisting up to above 100 K and displaying a broad transition as previously seen in the susceptibility data. No spin gap is observed in the dispersive spin excitations at low momentum transfer, which is consistent with the gapless spinon continuum expected from the coordinate Bethe ansatz. However, the temperature dependence of the excitation spectrum gives evidence of some very weak interchain coupling. 

\end{abstract}
\maketitle

\section{Introduction}
One-dimensional (1D) antiferromagnetic (AFM) spin-chain systems \cite{sutherland2004beautiful} have attracted considerable attention since the discovery of gapless spinon excitations originating from the coordinate Bethe ansatz \cite{Bethe_intro} in $S$ = 1/2 chain systems \cite{faddeev1981spin, PhysRev.128.2131} and topological Haldane gap phases in $S$ = 1 chain systems \cite{haldane1983continuum,PhysRevLett.50.1153}. Indeed, the variety of 1D magnets with low (quantum) spin continues to provide a reliable precision testbed for models which obey fractional exclusion statistics \cite{PhysRevLett.67.937}. As such, many important properties of these materials have been discovered and characterized, yet there is still an expanding arena of inquiry into the role of additional novel effects resulting from the large parameter space that results from the interplay between quantum spin, charge, and orbital degrees of freedom. The various magnetic phases that result from difference balances between these fundamental variables is still being actively explored. One such example is the family of spin chain systems with the formula A$_3$MM'O$_6$ (where A is an alkaline-earth metal Sr or Ca and M and M' are transition metals). This class of compounds exhibit many topical aspects of fundamental physics such as geometrical frustration, quantum criticality, and ferro-electricity \cite{lemmens2003magnetic}. There are many possible choices for the M and M' ions (magnetic and non-magnetic) in A$_3$MM'O$_6$, which have resulted in numerous investigations into their properties \cite{stitzer2001advances}.

The focus of the present paper will be limited to magnetic spectroscopy measurements for the compound Sr$_3$CuPtO$_6$ (SCPO) specified by the alkaline A$^{2+}$ = Sr$^{2+}$, the magnetic ion M$^{2+}$ = Cu$^{2+}$ (d$^9$, $S$ = 1/2), and the nonmagnetic ion M'$^{4+}$ = Pt$^{4+}$ (d$^6$, $S$ = 0). SCPO is a magnetic insulator composed of 1D chains which are arranged in an anisotropic triangular lattice formation when viewed perpendicular to the chain axis. This leads to the possibility of interchain coupling introducing frustration into the system, and this frustration would then be expected to have some non-trivial effects on the spin excitations. This was considered as a possible reason behind the observed absence of long-range magnetic ordering in the $S$ = 1 isostructural compound Sr$_3$NiPtO$_6$ (SNPO) \cite{PhysRevB.75.214422, SNPO_SL1_1}. However, current understanding points to strong anisotropies and spin-singlet states dominating over any Haldane phase in that case \cite{PhysRevB.82.094431, PhysRevLett.66.798}. 


It has been reported from bulk magnetic susceptibility and heat capacity measurements that the magnetic Cu$^{2+}$ ions in SCPO exhibit Heisenberg spin chain behavior. Both of these two physical properties are well described by a model of isotropic spin-half chains down to 5 K \cite{PRB_2004_heatcapacity}, i.e. there is a broad peak centered around 35 K, which is a strong indicator of the characteristic short range spin fluctuations of 1D magnetism. However, so far there has been considerable ambiguity in resolving how much interchain coupling is present in the system. This is mainly due to the uncertainties involved in fitting the susceptibility curves of 1D systems with and without substantial interchain coupling. Initial reports on this compound showed it was possible to model the magnetic susceptibility data with significant AFM interchain coupling (e.g., for the ratio $J/J'$ $\approx$ 3, where $J$ and $J'$ refer to the intrachain and interchain couplings, respectively) \cite{Claridge}. 

\begin{figure}
	\centering
	\includegraphics[width=1.02\columnwidth,clip]{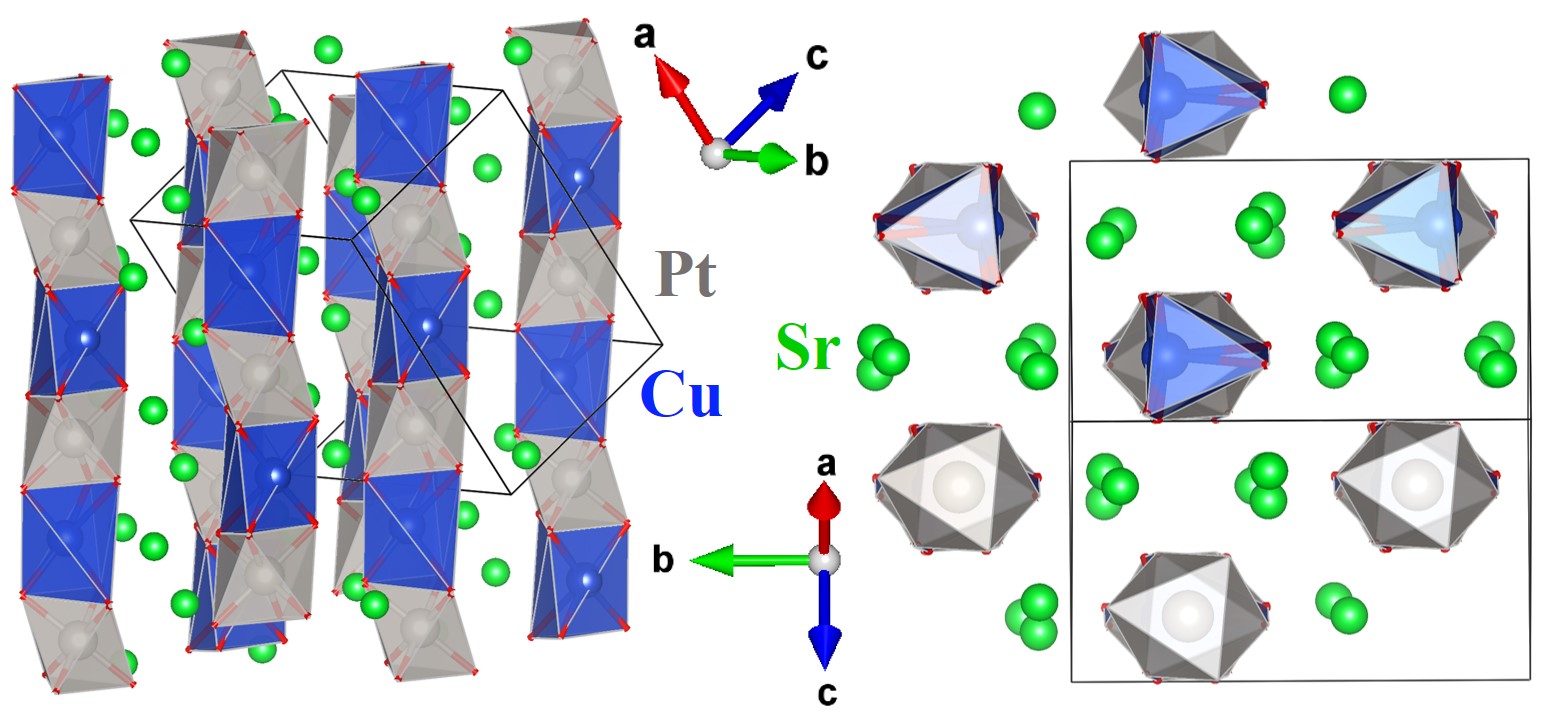}
	\caption{\label{structure} Crystal structure of Sr$_3$CuPtO$_6$, with the spin chains consisting of face-sharing CuO$_6$ trigonal prisms and PtO$_6$ octahedra running along the chain direction. When viewed along the chain direction, the chains form a triangular lattice. In each CuO$_6$ trigonal prism, the Cu$^{2+}$ ion is close to one ``square" face of the prism to achieve a ``square-planar" coordination environment.}
\end{figure}

Electronic structure calculations have also indicated that the intrachain interactions are dominant in SCPO \cite{electronic_calc_prb}, namely, $J$ $\approx$ 2.12 meV, $J'$ $\approx$ 0.65 meV, and $J/J'$ $\approx$ 3.2. The study also concluded that there should be only a very small magnetic anisotropy term (0.12 meV). Given that the interchain interaction $J'$ is found to be a significant fraction of the intrachain couplings $J$ by both computational and experimental methods, a case can be made that SCPO cannot be regarded as a true 1D antiferromagnet. 

Heat capacity measurements \cite{PRB_2004_heatcapacity}, which do not include contributions from broken chains or paramagnetic impurities, present quite a different picture. Below 5 K they show a certain deviation from a 1D spin chain model that could plausibly be attributed to the existence of a small excitation gap ($\Delta$ = 0.64 K $\sim$ 0.055 meV). And below 2 K there is a further deviation from the spin gap behavior, suggesting the onset of short range three-dimensional (3D) correlations. Taking 2 K as an upper bound for the N\'{e}el temperature T$_N$, it was estimated that $J/J'$ $\gtrsim$ 130, an indication that this system is in fact close to being an ideal 1D $S$ = 1/2 spin chain. However, this estimate implicitly assumed unfrustrated interchain coupling with the chains forming a square lattice. 

Spinon correlations are very fragile and thus the characteristics of ideal 1D spinon excitations can be easily disrupted by the introduction of even tiny amounts of interchain interactions. However, frustrated interchain interactions allow for spinons to tolerate larger interchain interaction strengths before they pass into the 2D realm. Given this current uncertainty as to the relative strength of the interchain couplings in SCPO, the goal of this work is to find new evidence that would allow for a firm evaluation of the degree to which SCPO is a 1D spin-chain system. It is also of interest to unambiguously determine the value of the magnetic exchange constants and also whether or not any gap in the magnetic excitation spectrum is present. 

\section{Crystal Structure and Synthesis}
\begin{figure}
	\centering
	\includegraphics[width=1.02\columnwidth,clip]{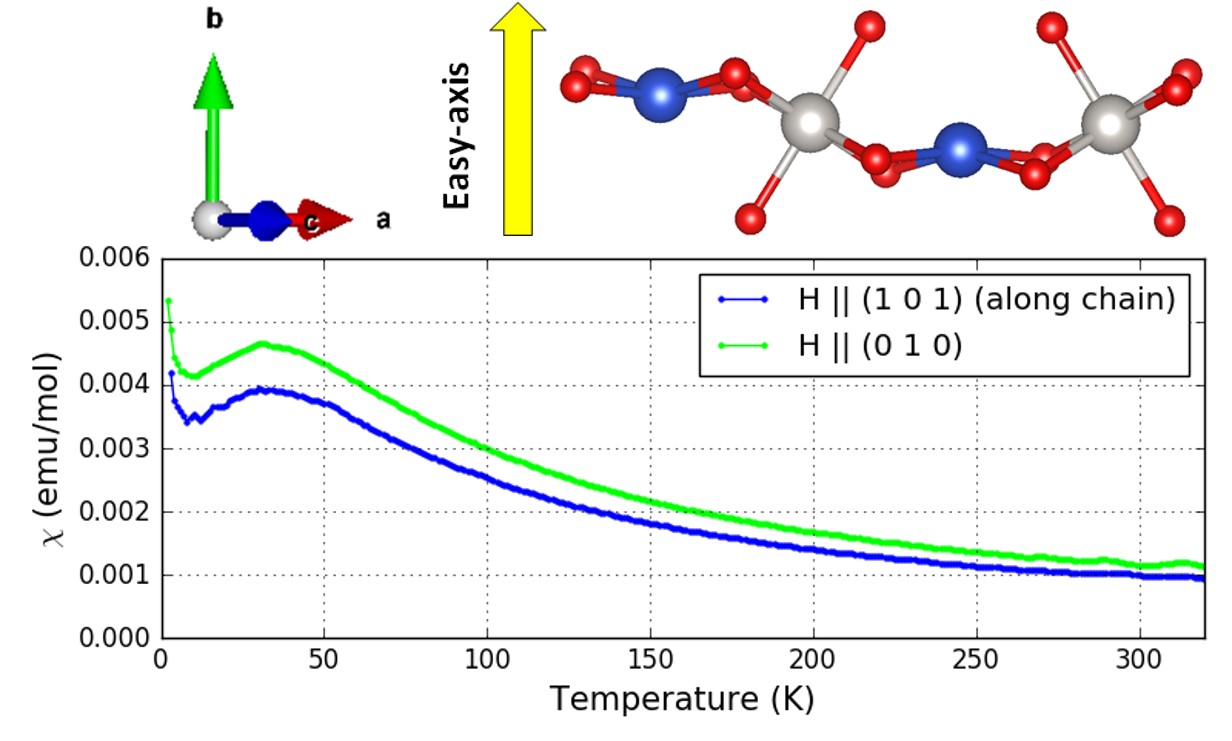}
	\caption{\label{SCPO_X} Magnetic susceptibility data for single crystals of Sr$_3$CuPtO$_6$ with an applied magnetic field of $\mu_0H$ = 2 kOe applied in the labeled orientations relative to the spin chains. The data were collected after zero field cooling. The chain structure shown may be compared with Fig. \ref{structure}.}
\end{figure}

The Sr$_3$MPtO$_6$ (M = Co, Ni, Zn) phases are isostructural, crystallizing in the rhombohedral space group R$\bar{3}$c \cite{mikhailova2014structure,nguyen1994synthesis,lampe1996structure}. However, the structure of SCPO is slightly different from that of Sr$_3$MPtO$_6$ (M = Co, Ni, Zn) in that each CuO$_6$ trigonal prism has its Cu$^{2+}$ ion located near the center of one ``square" face of the prism. The resulting ``CuO$_4$ square planar" units form a zigzag chain along the [1 0 1] direction (Figs \ref{structure} and \ref{SCPO_X}) \cite{hodeau1992structure}. The relative arrangements of the CuO$_4$ square planar units are identical in all the MPtO$_6$ chains, hence lowering the symmetry to the C2/c monoclinic space group with the following lattice parameters: a = 9.31(1), b = 9.72(1), and c = 6.68(1) $\mathrm{\AA}$ with $\beta$ = 91.95$^{\circ}$. Note that in this C2/c structure, the midpoint between every two nearest-neighbor Cu$^{2+}$ ions (i.e., the Pt$^{4+}$ ion site) of the 1D chain is an inversion center \cite{arioka1998esr}. Consequently, there is no Dzyaloshinskii-Moriya (DM) interaction in SCPO. 


Single crystals of SCPO can be synthesized via flux growth methods as described in Claridge \textit{et al.} \cite{Claridge}. However, reasonable quality single crystals can only be grown to masses on the order of milligrams with these recipes. Thus, unfortunately, it is not known how to grow single crystals of SCPO to the necessary size for use in the main experimental tool of this study: inelastic neutron scattering (INS). Fortunately, as will be shown, most of the necessary information on the spin excitations can be obtained from INS measurements with high quality polycrystalline samples. This is especially true for 1D materials, since the relatively simple powder integration procedure for a 1D excitation spectra may be more readily deconvoluted in order to extract key information that would otherwise be lost in the powder averaging \cite{oneDpowderextract}. For this study, a 10g polycrystalline sample of Sr$_3$CuPtO$_6$ was prepared by solid-state reactions of CuO, PtO$_2$, and SrCO$_3$. 

\section{Results}
\subsection{Magnetic Susceptibility}
Magnetic susceptibility data for the SCPO single crystal samples grown for this study is shown in Fig. \ref{SCPO_X}. These single crystal magnetic susceptibility data were measured on a SQUID magnetometer (Quantum Design) in an applied field of 0.2 T after zero field cooling. These measurements show behavior generally consistent with that presented in previous reports \cite{Claridge,PRB_2004_heatcapacity,chattopadhyay2013anisotropic}. The expected features such as the broad peak centered around 35 K are present. The Curie tail observed at low temperature is believed to result from broken chains and/or paramagnetic impurities and is not indicative of long-range order \cite{PRB_2004_heatcapacity}.

However, the direction-dependent susceptibility previously reported in Claridge \textit{et al.} \cite{Claridge} showed some qualitative differences in the T $<$ 20 K region. Our data show no such difference in the direction-dependent susceptibility curves, except for the overall magnitude of the susceptibility that depends on the orientation relative to the chains.  One possible reason which would explain this discrepancy is that samples used in this study contain impurity spins which are mostly Heisenberg type, whereas the samples of Claridge \textit{et al.} may have more impurity spins which are of Ising or XY type. Impurity spins originating from chain severing may have a different anisotropy than those originating from Cu$^{2+}$ impurities, and it is plausible that the two samples have different ratios of these two kinds of impurity spins. 

To gain further insight into these various results, we carried out density functional theory (DFT) calculations by employing the frozen-core projector augmented wave method \cite{WB20,WB21} encoded in the Vienna ab initio simulation package \cite{WB22,WB23}, and the generalized-gradient approximation of Perdew, Burke and Ernzerhof \cite{WB24} for the exchange-correlation functional. The electron correlation in Cu 3d states was taken into consideration in terms of the DFT+U method \cite{WB25} by adding the effective on-site repulsion U$_{\mathrm{eff}}$ on the Cu sites. Our DFT+U calculations including spin-orbit coupling (SOC) show that the spin orientation along the b-axis direction is more stable than that along the (a+c) direction by 1.15, 1.15 and 1.14 meV/Cu for U$_{\mathrm{eff}}$ = 4, 5, and 6 eV, respectively. That is, the preferred spin orientation is the b direction, i.e., perpendicular to the CuO$_4$ ``square plane", and hence perpendicular to the chain direction. The measured magnetic susceptibility shown in Fig. \ref{SCPO_X} is strongest along the directions perpendicular to the chain, which is consistent with these computational results. 

\begin{figure*}
	\centering
	\includegraphics[width=1.01\textwidth,clip]{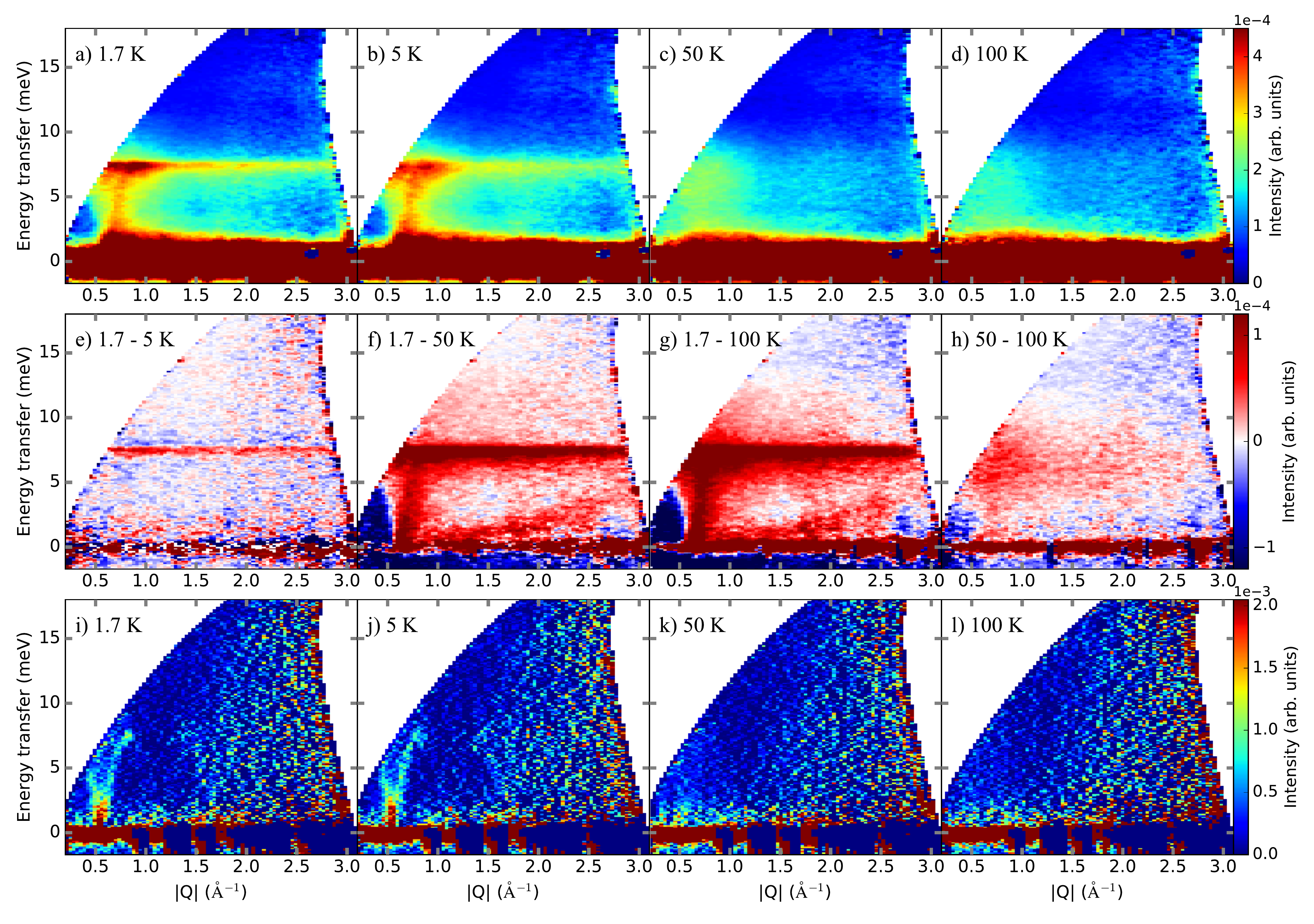}
	\caption{\label{data_22meV} Temperature dependence of the magnetic excitations in Sr$_3$CuPtO$_6$ (measured with E$_i$ = 22 meV). \\(a-d) Evolution of Q-E scattering intensity at the indicated temperatures with the empty sample-holder background subtracted. (e-h) The indicated temperature differences for comparison. (the faint parabolic line is from Helium recoil scattering) \\
	(i-l) 1D dispersion data extracted from the powder data in (a)-(d) using the conversion method of \textit{Tomiyasu et al.}\cite{oneDpowderextract}}. 
\end{figure*}

\begin{figure}
	\centering
	\includegraphics[width=1.02\columnwidth,clip]{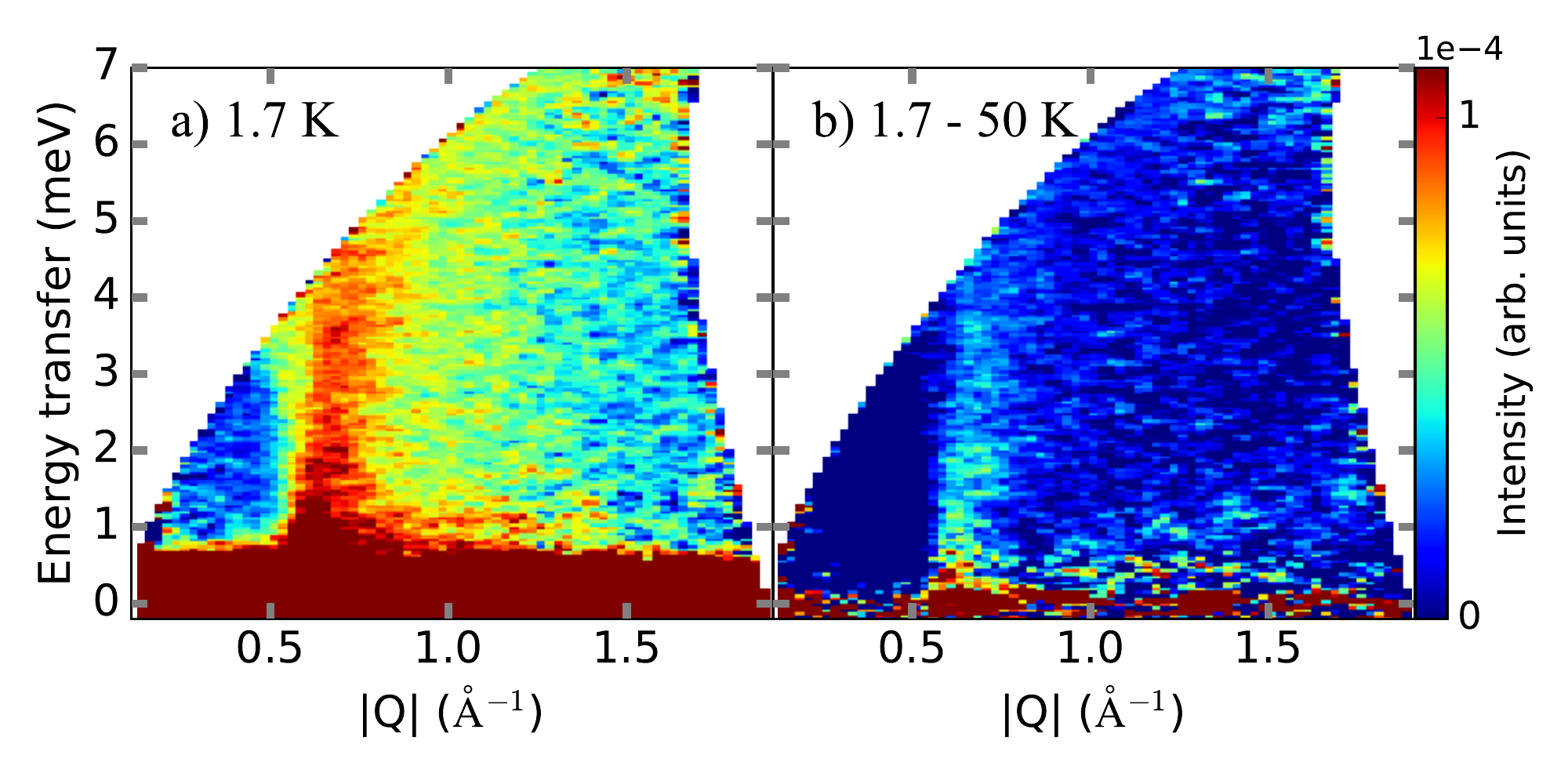}
	\caption{\label{data_8meV} INS for SCPO taken with E$_i$ = 8 meV. (a) Data collected at 1.7 K with the empty-can background subtracted. (b) Difference between the 1.7 K and 50 K data. No gap in the magnetic excitation spectrum at the zone center is observed. }
\end{figure}

\subsection{Magnetic Excitations}
\subsubsection{Description of the INS Data}
The magnetic excitation spectra for SCPO were obtained by inelastic neutron scattering (INS) measurements, which were carried out at the SEQUOIA Fine-Resolution Fermi-chopper Spectrometer at the Spallation Neutron Source at ORNL \cite{granroth2010sequoia, stone_review}. The polycrystalline sample was held within a cylindrical aluminum can with He-exchange gas and connected to a helium flow cryostat which could reach a base temperature of 1.7 K. The INS measurements for this sample were conducted at temperatures of 1.7, 5, 50, and 100 K in order to capture the full magnetic excitation spectra at each representative temperature along the features of the $\chi(T)$ and $C_{mag}(T)$ curves. For each temperature, incident neutron energies of E$_i$= 8 and 22 meV were used. For E$_i$=22 meV, the Fermi chopper frequency was set at 240 Hz which provided a full-width at half-maximum (FWHM) elastic energy resolution of $\delta E$ = 0.44 meV. For E$_i$=8 meV, the Fermi chopper frequency was set at 120 Hz which provided a FWHM elastic energy resolution of $\delta E$ = 0.16 meV. The empty sample-holder contributions to the background were subtracted from the data for each of the given conditions. 

Comparing the Q-E magnetic excitation spectrum for various temperatures as shown in Fig. \ref{data_22meV}, we can observe several things. The most striking trend in Fig. \ref{data_22meV}(a)-\ref{data_22meV}(d) is that the magnetic excitations broaden roughly in accordance with the $\chi(T)$ and $C_{mag}(T)$ curves. The magnetic excitations become more diffuse as the temperature is increased. And there is a qualitative transition in the overall spectra shape between 5 and 50 K, consistent with the broad peak in $\chi(T)$ being centered at 35 K. Now considering the spectra at low temperature, two main features of these well-defined and sharp excitations are apparent. One is the flat feature at 7.5 meV, which extends out to high momentum transfer and has an intensity that falls off proportionally to the Cu$^{2+}$ magnetic form factor. The other feature is the column of dispersive intensity emerging upwards from $Q$=0.59 $\mathrm{\AA}^{-1}$.  It can be seen in Fig. \ref{data_22meV}(e)-\ref{data_22meV}(h) that the flat feature gains intensity as the temperature is lowered to 5 K roughly in proportion to the dispersive feature. Below 5 K, however, only the flat feature gains intensity.

Making use of the conversion method described in Tomiyasu \textit{et al.} \cite{oneDpowderextract}, we can extract the 1D dispersion information directly from the poly-crystalline data presented in Fig. \ref{data_22meV}(a)-\ref{data_22meV}(d). The results of this conversion are shown in Fig. \ref{data_22meV}(i)-\ref{data_22meV}(l). For the 1.7 K and 5 K data at low momentum transfers, clear dispersion curves are observed which are consistent with the spinon structure factor model that we will describe in detail later in this section. As in the powder data, lowering the temperature from 5 to 1.7 K results in an intensity gain in the dispersion curve which appears to be concentrated around 7.5 meV in energy transfer. This interesting occurrence is also discussed in further detail later in this section. For the 50 K and 100 K cases, the extracted 1D structure factor appears to show only very faint diffuse intensity at low momentum transfers, consistent with a decreasing length scale of the spin correlations.

\subsubsection{Upper bound on the Spin Gap}
Shown in Fig. \ref{data_8meV} is the low temperature data taken at an incident energy of E$_i$=8 meV. The finer resolution allows for the determination of whether or not there is an observable gap in the excitations at the zone center; $Q$ = 0.59 $\mathrm{\AA}^{-1}$. As noted earlier, heat capacity data suggest the possibility of a spin gap \cite{PRB_2004_heatcapacity}, though quite small ($\sim$ 0.055 meV). In the case of an ideal uniform $S$ = 1/2 spin-chain system, a spin gap should not exist, but the distortions and irregularities in SCPO mean that one cannot rule out the possibility of a gap in the spin excitation spectrum. If there were some substantial inter-chain coupling present, as was postulated in Claridge \textit{et al.} \cite{Claridge}, this would lead to some dispersion of two spinon continuum along the plane perpendicular to the chain direction. This dispersion could in principle be observed in the powder-averaged $S(Q,E)$ as a hint of a gap opening in the dispersion minima at $Q$ = 0.59 $\mathrm{\AA}^{-1}$.

Fig. \ref{data_8meV}(a) shows no sign of any gap opening down to the elastic line resolution of 0.16 meV. Constant energy cuts along the 0 meV$<$ E$<$ 1 meV region also show no sign of a gap. This can be seen more clearly in Fig. \ref{data_8meV}(b), where the 50 K data is subtracted from the 1.7 K data, which to good approximation eliminates most of the intensity contribution from the elastic scattering and further constrains the magnitude of any possible gap. This leads to the conclusion that if there is any gap at the magnetic zone center, it must be smaller than 0.2 meV. This obviously does not rule out a very small gap of $\sim$ 0.055 meV, so this conclusion is still consistent with previous observations. Thus, this does confirm that this system is indeed extremely close to meeting the requirement of gapless magnetic excitations for $S$ = 1/2 systems.\\

\begin{figure*}
	\centering
	\includegraphics[width=1\textwidth,clip]{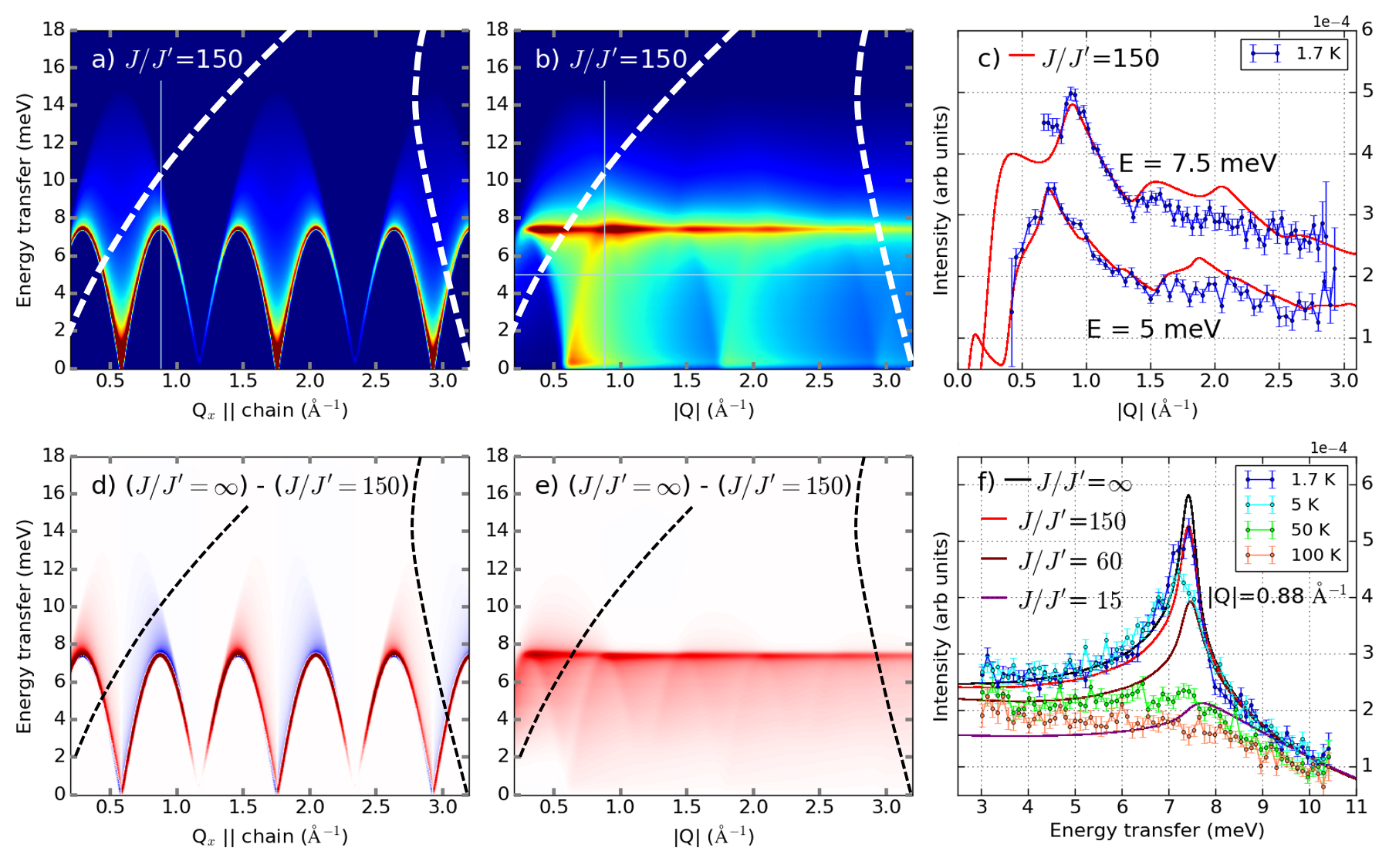}
	\caption{\label{powder_calc}(a) The $T=0$ dynamical structure factor calculated by treating the interchain coupling at the RPA level with the single chain susceptibility derived from 2+4 spinon structure factor from \textit{Caux and Hagemans}\cite{Caux_JSM} and the parameters given in Table \ref{couplings}. The form factor of the Cu$^{2+}$ ion has been taken into account. b) 1D powder average of this spinon structure factor. c) Comparison of the data at 1.7 K and theory using a constant energy cut at E = 5 meV (horizonal line in (b))(integrated between 4.8 - 5.2 meV) and E = 7.5 meV (integrated between 7.3 - 7.7 meV). The solid red lines are derived from theory described in the text with $J/J'$ = 150. d) The difference between the ideal 1D spinon structure factor and the 3D random phase approximation (RPA) treatment of that theory with an interchain coupling ratio of $J/J'$ = 150. e) Difference in the powder averages of the aforementioned structure factors. f) Comparison of data (at indicated temperatures) and theory using cuts at momentum transfer of Q = 0.88 $\mathrm{\AA}^{-1}$ (integrated between 0.85 - 0.91 $\mathrm{\AA}^{-1}$), shown by the straight vertical lines in (a),(b). The four solid curves are derived from theory  with the different $J/J'$ ratios indicated.}
\end{figure*} 

\subsubsection{Spinon Structure Factor Model}
We now address the general qualitative form of the powder-averaged INS cross-section $S(Q,E)$ excitation spectra at low temperature. First, it is necessary to point out the features of the spectrum that may serve as indicators of a 1D dispersion. One clue comes from the fact that the excitations dispersing upward from the elastic line at a wave-vector transfer of $Q$ = 0.59 $\mathrm{\AA}^{-1}$, which is the AFM $\pi$-point along the spin chain direction (1 0 1) where one reciprocal lattice unit is $2\pi$ = 1.1769 $\mathrm{\AA}^{-1}$. Furthermore, as $Q$ is increased and approaches this AFM point, the onset of magnetic scattering intensity happens in a sharp and abrupt manner, which is typically seen in powder averaged 1D systems. 

Indeed, as a first approximation, one can use powder averaged dispersion from linear spin wave theory (LSWT) with $J$ = 7.5 meV AFM intrachain coupling to get the correct bandwidth and the aforementioned features. However, this powder spectra from LSWT completely fails to capture the relative scattering intensity distribution across $Q-E$ space, as expected when quantum fluctuations are neglected in a $S$ = 1/2 1D system.  

Proper modeling for this system requires using a spinon dynamical structure factor \cite{PhysRevB.24.1429, PhysRevB.55.12510, PhysRevLett.111.137205}. Similar fitting of the spinon structure factor with powder INS data has been demonstrated in other cases \cite{oneDpowderextract,Nilsen2015}. For greatest accuracy, the spinon model should take into account excitations which produce 2 spinons as well as those which produce 4 spinons. The total intensity in the structure factor is almost all accounted for by both 2 spinon (73\%) and 4 spinon (26\%) excitations. The 2+4 spinon dynamical structure factor for 1D Heisenberg AFM $S$ = 1/2 spin chains has been calculated with great precision by \textit{Caux and Hagemans}.\cite{Caux_JSM} It is plotted in Fig. \ref{powder_calc}(a) (slightly modified by an RPA calculation as will be explained later) with the value for the intrachain $J$ given in Table \ref{couplings} and the momentum transfer scaled by 2$\pi$ = 1.1769$\mathrm{\AA}^{-1}$ in order to represent the $Q_x$ along the (1 0 1) chain direction. With these settings, the lower boundary of the spinon continuum is given by $\frac{\pi}{2}J|$sin$(Q_x)|$.  Also note that, importantly, the spinon structure factor model shown in Fig. \ref{powder_calc}(a) is entirely consistent with the extracted 1D structure factor shown in Fig. \ref{data_22meV}(i).

The powder average of the spectra in Fig. \ref{powder_calc}(a) is calculated following the procedure outlined in Tomiyasu \textit{et al.} \cite{oneDpowderextract}. The results are shown in Fig. \ref{powder_calc}(b). Evidently, this powder average of the 1D spinon excitation spectrum qualitatively reproduces the data in Fig. \ref{data_22meV}(a) with good fidelity. In particular, it shows an accurate relative distribution of intensities, and accounts for the finite intensity present above the dispersion maximum of 7.5 meV. The intrachain exchange coupling parameter $J$ = 4.73 meV models this bandwidth accurately, and is consistent with the lower range of experimental $J$ values obtained from previous characterizations. 

Fig. \ref{powder_calc}(c) shows constant energy transfer cuts at E = 5 meV and E = 7.5 meV through both the data (at 1.7 K) and theoretical 1D spinon structure factor (with $J/J'$ = 150, as described in detail in the \textit{RPA calculations} section). The overall excellent agreement between the data the theory allows us to expand on the implications based on this theoretical model with confidence. We note that the discrepancies in intensity which occur in the second Brillouin zone ($Q \approx$ 1.8 $\mathrm{\AA}^{-1}$) is likely due to the fact that the simple powder averaging procedure employed here neglects to account for the experimental statistical sampling of $Q$-points on the constant-$Q$ spheres that are being integrated.\\


\subsubsection{Temperature Dependence}
Fitting the INS spectra with only the intrachain exchange coupling appears to work reasonably well, immediately suggesting a very small $J'$. The origin of this large ratio of $J/J'$ must be due to the type and extent of wave-function overlap because the exchange path distances of $J'$ and $J$ are very similar as shown in Table \ref{couplings}. The details of this scenario were described by Majumdar \textit{et al.} \cite{PRB_2004_heatcapacity}. In each CuPtO$_6$ chain, the Cu$^{2+}$ (d$^9$, $S$ = 1/2) ion is magnetic whereas the Pt$^{4+}$ (d$^6$, $S$ = 0) is not. Since the Pt$^{4+}$ has unfilled 3d orbitals, it can be expected to participate in the intrachain exchange interaction $J$. The same cannot be said for the exchange pathway $J'$ through the  Sr$^{2+}$ in between the CuPtO$_6$ chains. 

As described earlier, the fitting of $C_{mag}(T)$ showed indications of interchain interactions opening up a gap below 5 K. Furthermore, below 2 K there is an additional anomalous deviation from the spin-gap model, which was speculated to be a signature of either 3D short-range magnetic ordering or a spin-Peierls-like transition \cite{PRB_2004_heatcapacity}. Also, this anomaly in $C_{mag}(T)$ conflicts with reported AC susceptibility measurements, which showed no such anomaly down to a temperature of 0.27 K on a polycrystalline sample of SCPO \cite{PhysRevB.61.11594}. Now with the data shown in Fig. \ref{data_22meV}(a), \ref{data_22meV}(b), and \ref{data_22meV}(e), it is possible to shed some light on this issue. 

\begin{table}
	\centering
	\begin{tabular*}{\columnwidth}{@{\extracolsep{\fill} } cccr }
		\hline\hline\noalign{\smallskip}
		$J_j$  & distance ($\mathrm{\AA}$) & $J$  \\ [0.5ex] 
		\hline\noalign{\smallskip}
		$J'$ (interchain) & 5.15 &  $\sim$0.032 meV\\ 
		$J$  ~(intrachain) & 5.71 &  4.73 meV\\
		
		\hline\hline
	\end{tabular*}
	\caption{ The two estimated magnetic exchange couplings of SCPO in order of their bond distance. These coupling values were used for the models shown in Fig. \ref{powder_calc}. }
	\label{couplings}
\end{table}

Fig. \ref{data_22meV}(e)-\ref{data_22meV}(h) 
show that the flat feature at 7.5 meV gets more intense as the temperature is lowered, as noted earlier. Moving from 5 to 1.7 K [Fig. \ref{data_22meV}(e)], the only significant change is some further intensity gain in this flat feature. This intensity gain at 7.5 meV can also be seen by comparing the extracted 1D structure factors in Fig. \ref{data_22meV}(i) and \ref{data_22meV}(j). An explanation for this behavior may come from a theoretical study by Kohno \textit{et al.} \cite{kohno2007spinons} which considers the spin excitation spectra of 2D triangular lattice compounds, such as Cs$_2$CuCl$_4$ \cite{PhysRevB.68.134424} and Cu(Y/La)$_2$Ge$_2$O$_8$ \cite{PhysRevB.95.144404}, with magnetic couplings close to the limit of 1D spin chains. Kohno \textit{et al.} showed that (FM or AFM) interchain interactions between the 1D spin chains introduces an (attractive or repulsive) force between spinons, resulting in delocalized composite particles called (bound or anti-bound) triplons which can move coherently between chains \cite{mckenzie2007quantum}. Spinon attaction leads to an increase in intensity at the lower edge of the spinon continuum and a downward shift in the spectral weight of the continuum. Spinon repulsion has the inverse effect: suppression of the spectral weight at the lower edge of the spinon continuum and an upward transfer of spectral weight in the continuum. Therefore, the flat feature in the \textit{temperature difference} plot of Fig. \ref{data_22meV}(e) could originate from an intensity change due to some finite amount of FM or AFM interchain coupling becoming strong enough to have an effect. The constant-Q cuts at the BZ zone boundary (Q=0.88 $\mathrm{\AA}^{-1}$) in Fig. \ref{powder_calc}(f) show more precisely the relative change in intensity as the temperature is lowered from 5 to 1.7 K. This should serve as confirmation that weak interchain coupling does indeed manifest itself in the low temperature ground state of SCPO, though its effect is quite subtle and consistent with the ratio of $J/J'> 130$ as estimated previously \cite{PRB_2004_heatcapacity}. \\

\subsubsection{RPA calculations}\label{rpa}
To put these assertions on more solid footing, we have employed a random phase approximation (RPA) treatment of the interchain coupling. This approach \cite{PhysRevB.64.094425} predicts that 
\begin{equation}
    \chi(\omega,{\bm k})=\frac{\chi_\mathrm{1D}(\omega,k)}{1-2J^\prime({\bm k})\chi_\mathrm{1D}(\omega,k)}
\end{equation}
where $\chi_\mathrm{1D}$ is the dynamic magnetic susceptibility for a one-dimensional spin chain, $J^{\prime}({\bm k})$ is the Fourier transform of the interchain coupling, and $k$ is the component of $\bm k$ along the chain direction. Importantly, it was noted by Kohno \textit{et al.} \cite{kohno2007spinons} that the structure factor obtained by their methods is in close agreement with that obtained using the RPA method. This method has been applied several times to reliably explain experimental data on similar systems where $J/J' > 1$ \cite{PhysRevLett.104.237207, PhysRevLett.119.087204}.

We treat interchain coupling to the six nearest neighbour chains and assume an AFM zigzag interchain coupling due to the offset of the chains (see Fig. \ref{structure}). Thus, the model is an 3D hexagonal analogue of model in the planes of Cs$_2$CuCl$_4$ and $\kappa$-(BEDT-TTF)$_2X$ \cite{PhysRevB.64.094425,PhysRevB.68.134424,PhysRevLett.119.087204}.  This model neglects the distortion of the chains, which will lift the frustration somewhat in the real material. Using the aforementioned zero-temperature 1D dynamical structure factor from Caux and Hagemans \cite{Caux_JSM}, we find $\chi_\mathrm{1D}$ using $S_{1D}(\omega,k)=-\mathrm{Im}[\chi_\mathrm{1D}(\omega,k)]$ and obtained $\mathrm{Re}[\chi_\mathrm{1D}(\omega,k)]$ via a Kramers-Kronig transformation. The structure factor calculated in this way for the parameters given in Table \ref{couplings} is shown in Fig. \ref{powder_calc}(a) along the path $(k/2,k/2,k)$, where one finds $J^\prime(k)=3J^\prime\cos(k/2)$, with the strongest renormalization of the structure factor due to interchain correlations. 

It is interesting to note that if we assume the upper bound to be $T_N \lesssim 2$ K, this RPA method yields $J/J' \gtrsim 3.9$, in contrast to the upper bound $J/J' \gtrsim 130$ found using an unfrustrated model \cite{PRB_2004_heatcapacity}. However, when the RPA treatment is applied with $J/J' \sim 3.9$, the resulting structure factor diverges hopelessly away from what we see in the experiment, as can be seen in Fig. \ref{powder_calc}(f). Thus, it is clear that according this RPA treatment only a large value of $J/J'$ will be consistent with the data. 

When we apply the RPA treatment with a value of $J/J'=150$, we see a deviation from the ideal 1D spinon structure factor consistent with the INS observations, which validates our aforementioned assertion. Specifically, in Fig. \ref{powder_calc}(f), the theory cuts at Q = 0.88 $\mathrm{\AA}^{-1}$ through the ideal 1D and 3D RPA calculated powder spectra differ from each other by the same magnitude as differences between the 1.7 and 5 K cuts through the data. 
Moreover, when the difference is taken between the purely one-dimensional and three-dimensional (RPA) powder averaged spinon structure factors with $J/J'=150$, as shown in Fig. \ref{powder_calc}(e), we find that it is consistent with the INS result shown in Fig. \ref{data_22meV}(e). In addition, we note that a comparison of of the ideal 1D and RPA calculations for $J/J'=150$, as shown in Fig. \ref{powder_calc}(d), show that finite $J'$ results in an upward shift in spectral weight at Q = 0.88 $\mathrm{\AA}^{-1}$, as predicted by Kohno \textit{et al.} \cite{kohno2007spinons}. However, it was not possible to definitively resolve this upward shift in spectral weight based on the corresponding 1D structure factors extracted from the data as shown in Fig. \ref{data_22meV}(i) and \ref{data_22meV}(j). 

These calculations neglect the effects of finite temperatures which would explain the peak intensity mismatch between the ideal 1D model and the 5 K data. However, taken as a whole, these  comparisons between theory and experiment show clearly that very small interchain coupling sufficiently explains the changes in the structure factor at low temperatures. Furthermore, the INS provides a much more stringent lower bound on the interchain coupling,  $J/J' \gtrsim 150$, than the absence of long-range magnetic order until at least 2~K, which only yields $J/J'>3.9$. \\

\section{Conclusion}
INS was employed to investigate polycrystalline SCPO, an experimental realization of a 1D quantum spin chain compound. Examination of the $S(Q,E)$ spectra for SCPO reveals a spinon excitation spectrum, which persists up to above 100 K. We note that this is well above the temperature ($\sim2$~K) at which short-range interchain correlations become important, indicating their 1D nature. Despite the use of a powder sample of SCPO, accurate modeling has been achieved by employing the 2+4 spinon dynamical structure factor for $S$ = 1/2 Heisenberg AFM spin chains \cite{Caux_JSM} with the interchain coupling treated at the RPA level \cite{PhysRevB.64.094425}. No spin gap is observed in the dispersive spin excitations at low momentum transfer, which is also consistent with the gapless spinon continuum expected from the coordinate Bethe ansatz. However, the temperature dependence of the excitation spectrum gives evidence of some interchain coupling being present, but at a much weaker magnitude then was postulated from some previous results.

\section{Acknowledgements}
We are grateful to J.-S. Caux for providing the spinon dynamical structure factor data. We also wish to thank D.T. Adroja and T. Saha-Dasgupta for helpful discussions. Work at the IBS (Institute for Basic Science) CCES (Center for Correlated Electron Systems) (South Korea) was supported by the research program of the Institute for Basic Science (IBS-R009-G1). Work at Rutgers University was supported by the DOE under Grant No. DOE: DE-FG02-07ER46382. Work at the University of Queensland was supported by the Australian Research Council through grants FT130100161 and DP160100060. This research used resources at the Spallation Neutron Source, a DOE Office of Science User Facility operated by the Oak Ridge National Laboratory (ORNL).

\bibliography{SCPO_ref.bib}
\end{document}